\documentclass[aps,prb,twocolumn,groupedaddress]{revtex4-1}
\usepackage{graphics}
\usepackage{graphicx}
\usepackage{epstopdf}
\usepackage{braket}
\usepackage{bm}
\usepackage{amsmath}
\usepackage{caption}
\usepackage{subcaption}
\usepackage{caption}

\begin{document}

\title{Determination of superexchange interactions for the CuO$_2$ chains in LiCu$_2$O$_2$}

\author{N. Aucar Boidi}
\email[]{nair.aucar@cab.cnea.gov.ar}

\affiliation{Centro At\'{o}mico Bariloche and Instituto Balseiro, 8400 Bariloche, Argentina}

\author{Christian Helman}
\affiliation{Centro At\'omico Bariloche and Instituto Balseiro, CNEA, 8400 S.
C. de Bariloche, Argentina}

\author{Y. N\'u\~nez-Fern\'andez}
\affiliation{CEA-Grenoble, IRIG-PHELIQS, 38000 Grenoble, France}

\author{K. Hallberg}
\affiliation{Instituto de Nanociencia y Nanotecnolog\'{\i}a CNEA-CONICET,
Centro At\'{o}mico Bariloche and Instituto Balseiro, 8400 Bariloche, Argentina}

\author{A. A. Aligia}
\affiliation{Instituto de Nanociencia y Nanotecnolog\'{\i}a CNEA-CONICET,
Centro At\'{o}mico Bariloche and Instituto Balseiro, 8400 Bariloche, Argentina}

\date{\today}

\begin{abstract}

Starting from \textit{ab-initio} calculations, we derive a five-band Hubbard model to describe the CuO$_2$ chains of LiCu$_2$O$_2$.
This model is further simplified to a low-energy effective Heisenberg model with nearest-neighbor (NN) $J_1$, and next-nearest-neighbor (NNN) $J_2$ interactions, combining perturbation theory, exact diagonalization calculations and Density Matrix Renormalization Group results. For realistic parameters we find the corresponding values of these interactions. The obtained effective model is consistent with a spiral-magnetic ground state as experimentally observed. 
Using symmetry arguments, the spiral state is a sufficient condition for the ferroelectricity observed in the system. 

\end{abstract}


\maketitle

\section{Introduction}
\label{intro}

Ferroelectric and magnetic materials have led to some of the most important technological advances to date. Ferroelectricity and magnetism combine in rather unusual materials called multiferroic, which offer the possibility to control the polarization by magnetic means and are expected to have technological applications \cite{kimu,hur,lore,goto,hur2,park,fiebig,hassanpour}.
In noncollinear magnets, one expects, on general symmetry arguments, a 
contribution to the electric polarization 
${\bf P} \sim {\bf e} \times {\bf S_1} \times {\bf S_2}$ where ${\bf e}$ is a unit vector connecting sites 1 and 2 and ${\bf S}_i$, $i=1,2$ are the spins on each site \cite{Katsura,mostovoy}.
Therefore, the existence of spiral magnetic order in chains is naturally expected to lead to
ferroelectricity. 
Recently, the relevance of noncollinear antiferromagnets for low-field spin caloritronics and magnonics has been stressed \cite{xu}.
In addition, the presence of a magnetic field is expected to lead to a considerable effect in the electric polarization of the compound \cite{kimu,eerenstein}.

This is the case of the quantum quasi-one-dimensional compound LiCu$_2$O$_2$. There is experimental evidence of the emergence of ferroelectricity when the spiral-magnetic state of the spin 1/2 chains sets in \cite{park}. Therefore, the system belongs to type-II multiferroics. The magnetic structures of the chains have been investigated by several experimental techniques \cite{masuda,masuda2,gip,miha,huang,yasui,koba,qi,bush} 
and \textit{ab-initio} calculations
\cite{gip}. In spite of some uncertainty in the parameters, the spiral order
is believed to result from the competition of nearest-neighbor (NN) $J_1$ and 
next-nearest-neighbor (NNN) $J_2$ interactions in a spin-1/2 Heisenberg chain.
Classically, a spiral magnetic order takes place when $J_2>0$ (antiferromagnetic) and $|J_1| < 4J_2$ \cite{bur,white}. The quantum cases have been studied by numerical \cite{bur,white,ess} and field-theoretical \cite{white,ner,allen,cabra}
techniques and the 
spiral order is confirmed.

Nevertheless, a justification of the microscopic model is lacking. The 
\textit{ab-initio} calculations \cite{gip,moser} fail to reproduce the 
experimentally observed charge gap \cite{papa}.
In addition, angle-resolved photoemission (ARPES) and optical measurements in single crystals of  LiCu$_2$O$_2$ \cite{papa} show features that cannot be reproduced by the existing calculations of the electronic structure and point to the need to resort to strongly correlated models.

In this work we study the electronic structure of the CuO$_2$ chains and derive
an effective Heisenberg model for them, taking into account the strongly correlated nature of the system, using a combination of different techniques.
Using hopping matrix elements obtained with maximally localized Wannier functions derived from \textit{ab-initio} calculations, on-site Coulomb repulsions 
typical of the superconducting cuprates, and atomic values for the exchange energy of O 2p orbitals, we derive a five-band Hamiltonian 
for the CuO$_2$ chains, with one orbital per Cu atom and two orbitals per O atom. Using perturbation theory in the hopping matrix elements, we derive the effective Heisenberg model. 

While this calculation sheds light on the different processes involved, due to the covalent
nature of the compound, the perturbative results for the
exchange interactions $J_1$ and $J_2$ are not accurate. 
Therefore, the values of these interactions are obtained by fitting the energies of the Heisenberg model to those
of a multiband Hubbard model for a CuO chain with eight unit cells, for each wave vector
and total spin projection.

The paper is organized as follows. In Section II we use first-principle calculations to determine the relevant orbitals and physical parameters of the CuO$_2$ chains and obtain the effective five-band Hubbard model, which includes local Coulomb interactions. In Section III, by taking into account the experimental filling of one hole per unit cell, we perform a low-energy reduction of this model to one in which the degrees of freedom are the spins 1/2 at the Cu sites, giving rise to a Heisenberg model with NN and NNN exchange interactions. These interactions are calculated by perturbation theory in the hopping matrix elements. As this turns out to be inaccurate for realistic values of the parameters, in Section IV we obtain the Heisenberg parameters by fitting the lowest energy levels of a multiband Hubbard model, which we calculate numerically using exact diagonalization. For the determination of the value of the 
difference between on-site energies of holes at O and Cu sites 
$\Delta $ we solve the five-band Hubbard model using the Density Matrix Renormalization Group (DMRG)\cite{dmrg}.
We present a summary and conclusions in Section V.

\section{Derivation of the five-band Hubbard model}
\label{5b}
The atomic structure of LiCu$_2$O$_2$, neglecting the spin spiral symmetry, is orthorhombic with space group $Pnma$, where the short side is in the CuO$_2$ chain direction ($a=5.72$\AA, $b=2.86$\AA, $c=12.40$\AA).
In the unit cell, the atoms are arranged in six planes perpendicular to the $c$-axis, where three are a mirror image of the others, with an offset in $a$ and $b$-crystal directions of 1.52\AA  
and 1.44\AA, respectively \cite{moser}.

\begin{figure}[h!]
\centering
\begin{subfigure}{0.24\textwidth}
    \caption{}
    \includegraphics[width=1\textwidth]{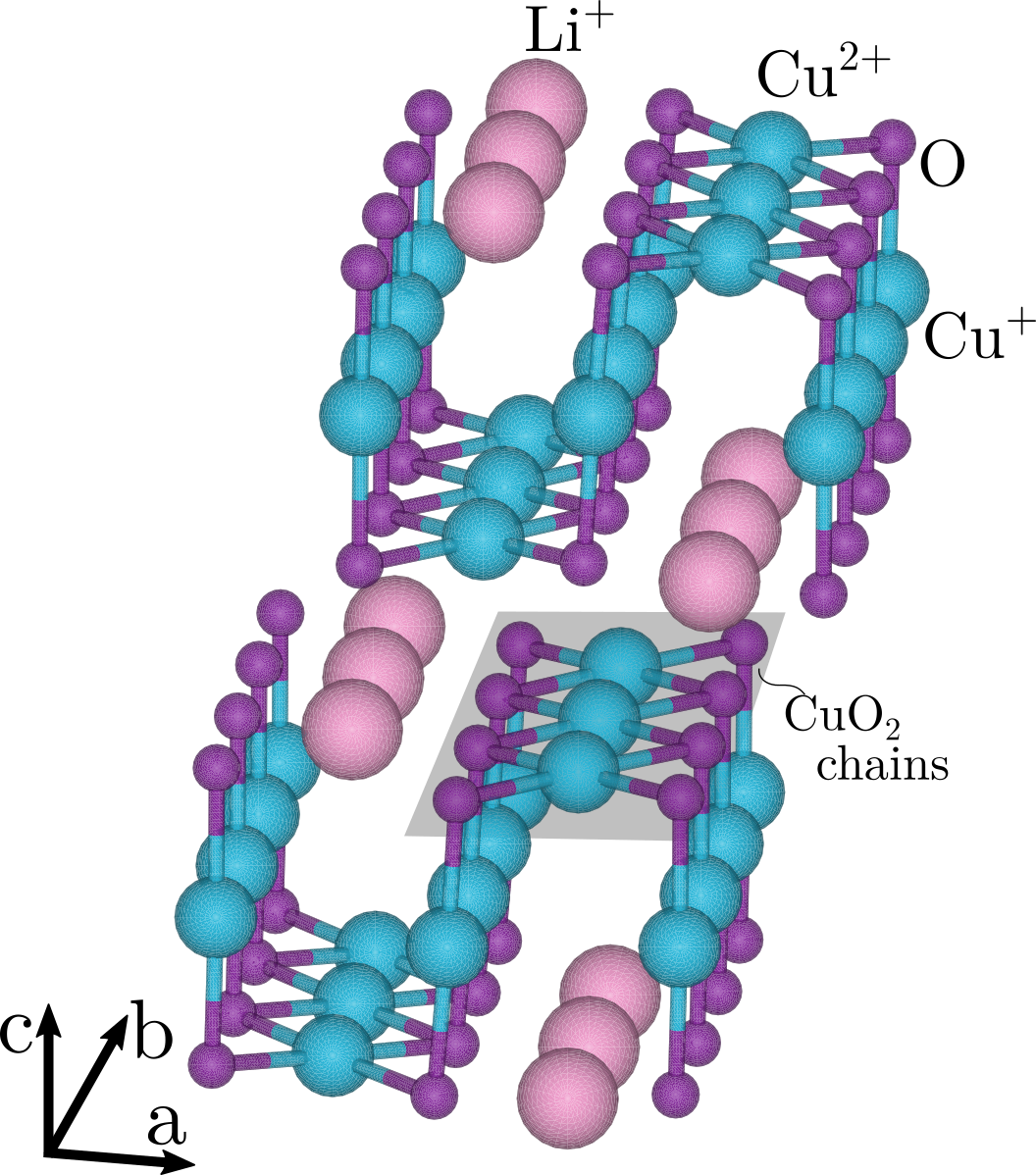}
    \label{fig:struct}
\end{subfigure}
\begin{subfigure}{0.23\textwidth}
    \caption{}
    \includegraphics[width=1\textwidth]{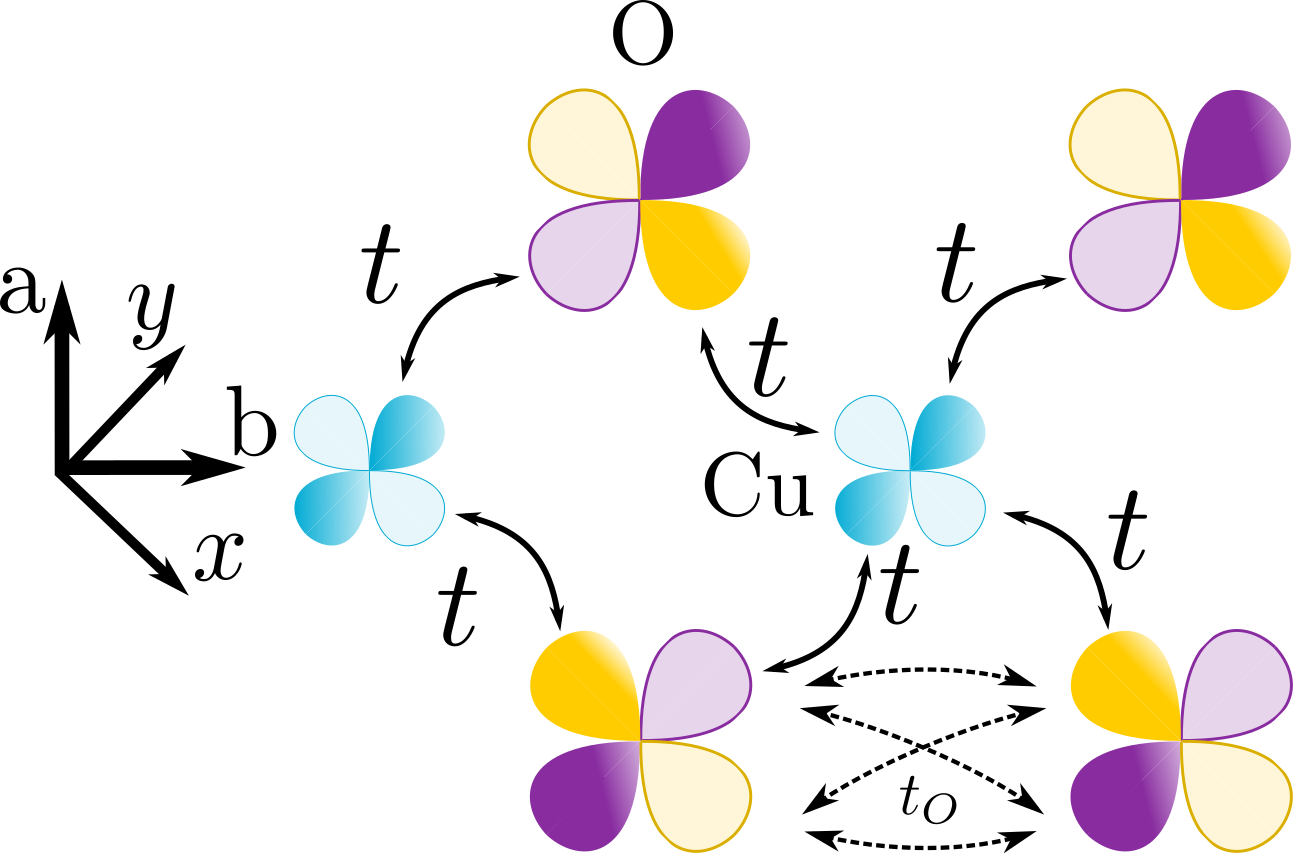}
    \label{system}
\end{subfigure}

\begin{subfigure}{0.47\textwidth}
    \caption{}
    \includegraphics[width=0.9\textwidth]{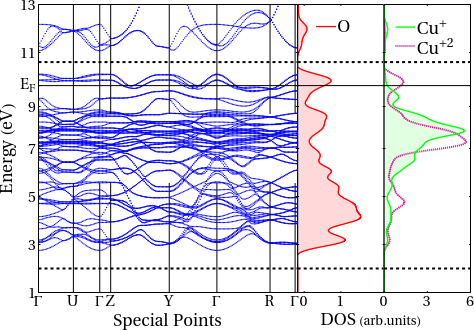}
    \label{energywindow}
\end{subfigure}
\caption{ (a) Crystal structure of the LiCu$_2$O$_2$. In green/blue/red the atoms of Li, Cu and O respectively. The gray areas indicate the Cu$^+$ planes that separate the Cu$^{2+}$O$_2$ chains. The orange plane indicates the natural cleavage section between the chains.
(b) (Color online) Scheme of the relevant orbitals and hoppings of the CuO$_2$ chains. Blue: $d_{xy}$, yellow: $p_x$, violet: $p_y$. The positive (negative)
part of the orbitals has darker (lighter) colors.
(c) Left: spin-unpolarized bands resulting from the DFT calculation. Dotted lines indicate the energy window where the MLWF method takes place. Right: density of states projected on each atom type. Note the difference between the Cu$^{2+}$ that belong to the chain, and the Cu$^+$ that belong to the plane}
\label{systema}
\end{figure}

As is well known, there are two types of Cu atoms\cite{bush,huang}; one type is a non-magnetic monovalent cation (Cu$^+$), while the other type is a magnetic divalent cation (Cu$^{2+}$).
The Cu$^+$ are located on planes perpendicular to the $c$-direction separating the Cu$^{2+}$ that form the CuO$_2$ chains in the $b$-direction.
In addition, these CuO$_2$ chains are separated in the $a$-direction by chains of Li$^+$ ions, as can be seen in Fig \ref{fig:struct}.
As a consequence of this distribution, a natural cleavage plane exists between the CuO$_2$ chains, implying a relatively weak bonding between them.
In the chain, the Cu$^{2+}$ atoms have four O nearest neighbors, forming a Cu-O-Cu angle of 94 degrees.
This square of O atoms around the Cu$^{2+}$ ions is where all the dominant magnetic interactions occur.
Since we focus on the magnetic chains, the Cu$^+$ lose relevance in our model and, henceforth, we will refer to the Cu$^{2+}$ ions as the Cu atoms.

In order to have an insight of the magnetic interactions in the chain, we perform \textit{ab-initio} calculations 
within density-functional theory (DFT) 
by means of the Quantum Espresso code \cite{QE}.
We use PAW pseudopotential with an energy cut of 80 eV for the Bloch wave functions along with GGA as the chosen exchange-correlation potential and the mesh in reciprocal space is 16x32x7.
As expected, the relevant orbitals for the electronic structure of the CuO$_{2}$ chains
are the $3d$ orbitals of Cu that point towards their NN O atoms and the two $2p$ orbitals of O that point towards their NN Cu atoms. 
We choose the direction of the chain in terms of unit vectors as
$\mathbf{\hat{x}}+\mathbf{\hat{y}}$, 
so that these relevant orbitals have symmetries $d_{xy}$ and 
$p_{x}$, $p_{y}$. See Fig. \ref{system}.
Density of state analysis (not shown) of a collinear magnetic configuration shows a strong hybridization between Cu-$d_{xy}$ and O-$p_x,p_y$ orbitals. 
The magnetic moments of the atoms in the chain are 0.53$\mu_B$ and  0.19$\mu_B$ for Cu and O atoms, respectively.

The Hamiltonian takes the form

\begin{equation}
H=H_{0}+H_{\text{hop}},  \label{htot}
\end{equation}
where $H_{0}$ contains the difference $\Delta $ between on-site energies of
holes at O and Cu sites and the interaction terms:

\begin{eqnarray}
H_{0} &=&\sum_{j\alpha \sigma }\Delta n_{j\alpha \sigma }
+U_{\text{Cu}}\sum_{i}n_{i\uparrow }n_{i\downarrow }+U_{\text{O}}\sum_{j\alpha
}n_{j\alpha \uparrow }n_{j\alpha \downarrow }  \notag \\
&&+(U_{\text{O}}-3J_{\text{O}})\sum_{j\sigma }n_{jx\sigma }n_{jy\sigma }
-J_{\text{O}}\sum_{j\sigma }p_{jx\sigma }^{\dagger }p_{jy\bar{\sigma}}^{\dagger
}p_{jy\sigma }p_{jx\bar{\sigma}}  \notag \\
&&+J_{\text{O}}\sum_{j}\left( p_{jx\uparrow }^{\dagger }p_{jx\downarrow
}^{\dagger }p_{jy\downarrow }p_{jy\uparrow }+\text{H.c.}\right) ,  \label{h0}
\end{eqnarray}
where the sum over $i$ ($j$) runs over all Cu (O) sites, $n_{j\alpha \sigma
}=p_{j\alpha \sigma }^{\dagger }p_{j\alpha \sigma }$, where $p_{j\alpha
\sigma }^{\dagger }$ creates a hole on the $p_{\alpha }$ 
($\alpha =x$ or $y$) orbital of O with spin $\sigma $, and $n_{i\sigma }=d_{i\sigma }^{\dagger
}d_{i\sigma }$, where $d_{i\sigma }^{\dagger }$ creates a hole on the $d_{xy}
$ orbital of Cu at site $i$ with spin $\sigma $. The Coulomb interaction
between two holes in the Cu (O) ions is $U_{\text{Cu}}$ ($U_{\text{O}}$)
and $J_{\text{O}}$ is the Hund's coupling between oxygen orbitals,
as well as the hopping between singlet pairs [last term of Eq. (\ref{h0})]. 

The O part of the interaction has been calculated in terms of two parameters
$U_{\text{O}}=F_{0}+4F_{2}$ and $J_{\text{O}}=3F_{2}$ in a similar way as for $d$
electrons \cite{spli}. We determined the value of $F_{2}=0.279$ eV from a
fit of the atomic energy levels of neutral O \cite{moore}. 
We have taken $U_{\text{Cu}}=10$ eV and $U_{\text{O}}=3$ eV from typical values in the
cuprates \cite{hybe} but also study the dependence of the results with 
$U_{\text{O}}$, since it affects the results in a sensitive way.

The hopping matrix elements have the following form

\begin{eqnarray}
H_{\text{hop}} &=&t\sum_{is\alpha }(p_{i+sR\mathbf{\hat{x}},x\sigma
}^{\dagger }+p_{i+sR\mathbf{\hat{y}},y\sigma }^{\dagger })d_{i\sigma } 
\notag \\
&&-t_{\text{O}}\sum_{j\beta \alpha \sigma }p_{j+R(\mathbf{\hat{x}}
+\mathbf{\hat{y}}),\beta \sigma }^{\dagger }p_{j\alpha \sigma }+\text{H.c.,}
\label{hop}
\end{eqnarray}
where $s=\pm 1$ and $R$ is the Cu-O distance in the chains.  The phase of the $p_{j\alpha }$ orbitals
displaced from the Cu in the $\mathbf{\hat{x}}$ or $-\mathbf{\hat{y}}$
directions (below the Cu-Cu line, see Fig. \ref{system}) has been changed by
a factor -1 so that the Cu-O hopping has the same sign in all directions.

For the hopping matrix elements we follow a two step procedure.
First, we perform a spin-unpolarized DFT calculation where the Bloch states are obtained.
The band structure is shown in Fig. \ref{energywindow}, where the O and Cu atoms of the chain share a peak at the Fermi level, suggesting a strong hybridization between them.

In order to have an atomic-like description, our second step is to use the method proposed by Marzari and Vanderbilt \cite{Vanderbilt} to find the Wannier functions in real space.
This method consists in projecting the Bloch states into a set of Wannier functions that are maximally localized in the target atoms.

The wannierization process needs to define an energy window where the projections take place.
In Fig. \ref{energywindow}, the energy window is
enclosed within the dotted lines.
As it is evident from the figure, it is not necessary to perform a disentanglement
procedure because the bands have no overlap with other bands outside the defined region.
From an analysis of the projected density of states (shown in Fig. \ref{energywindow}), we found that the main contribution to the bands in that energy window comes from
$p$ orbitals of O and $d$ orbitals from Cu.
Furthermore, the obtained Wannier Hamiltonian does not show appreciable contribution to the Cu-O hopping from the Cu that lies outside the spin chain.
This fact supports our model that only takes into account the Cu-O hopping in the chains.
The obtained Wannier basis is well localized in the atoms of the spin chain and the wave functions respect the $p$ and $d$ symmetry for each atom case.

Changing the basis to that used in Eqs. (\ref{h0}) and (\ref{hop}) we obtain  $t=1.1$eV, 
$t_{\text{O}}=0.36$eV.

\section{Effective Heisenberg model and perturbation theory}

\label{pert}

Since the occupancy of the CuO$_{2}$ chains is 1 per unit cell, in the limit
in which $t,t_{\text{O}}\ll \Delta $ the only relevant degrees of freedom
are the spins 1/2 at each Cu site. The hopping terms lead to effective
Heisenberg interactions between two spins. Collecting the most relevant
terms, the low-energy effective Hamiltonian takes the form

\begin{equation}
H_{\text{Heis}}=J_{1}\sum_{\braket{i,j}}\bm{S}_{i}\cdot \bm{S}_{j}+J_{2}\sum_{\braket{\braket{i,j}}}\bm{S}_{i}\cdot \bm{S}_{j}
\label{eqn:heis}
\end{equation}
where $\displaystyle\braket{i,j}$ indicates NN and 
$\displaystyle\braket{\braket{i,j}}$ NNN. The contribution to order $N$ in $H_{\text{hop}}$
to the effective matrix element between two states $|g_{1}\rangle $ and 
$|g_{2}\rangle $, which are part of the degenerate ground state of $H_{0}$
with energy $E_{g}$, is given by

\begin{equation}
\Delta E=-\sum_{e_{i}}\frac{\langle g_{1}|H_{\text{hop}}|e_{1}\rangle
\langle e_{1}|H_{\text{hop}}|e_{2}\rangle ...\langle e_{N-1}|
H_{\text{hop}}|g_{2}\rangle }{\left( E_{1}-E_{g}\right) \left( E_{2}-E_{g}\right)
...\left( E_{N-1}-E_{g}\right) },  \label{correc}
\end{equation}
where the sum runs over all intermediate states of $H_{0}$ $|e_{1}\rangle
...|e_{N-1}\rangle $ with energies $E_{1}...E_{N-1}.$

\begin{figure}[h!]
\centering
\includegraphics[width=0.25\textwidth]{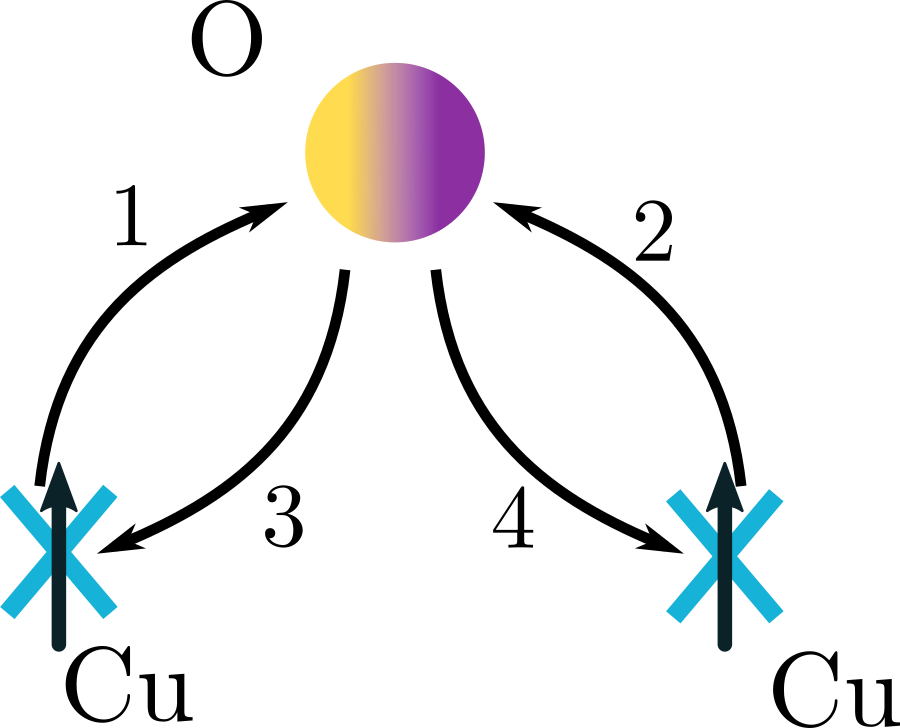}  
\caption{(Color online) Scheme of the ferromagnetic contributions to $J_1$}
\label{pert1}
\end{figure}

\begin{figure}[h!]
\centering
\includegraphics[width=0.3\textwidth]{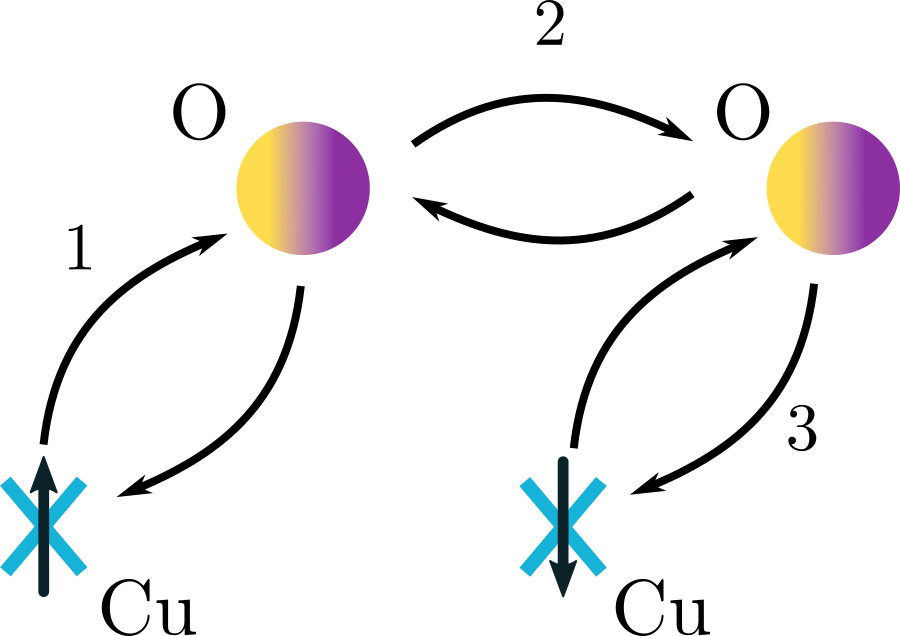}  
\caption{(Color online) Scheme of the antiferromagnetic contributions to $J_1$}
\label{pert2}
\end{figure}

The main perturbative contributions to the exchange interactions are (a) a
ferromagnetic contribution to $J_{1}$, approximately of order
$t^{4} J_{\text{O}}/[\Delta ^{2}(\Delta +U_{\text{O}})^{2}]$ (see Fig. \ref{pert1}) and
(b) antiferromagnetic contributions to both $J_{1}$ and $J_{2}$ (the former
being two times larger) roughly of order $t^{4}t_{\text{O}}^{2}/[\Delta
^{4}(\Delta +U_{\text{O}})]$ (see Fig. \ref{pert2}). 

In the first case (a), the holes of two NN Cu ions jump to 
different $2p$ orbitals of the same O atom, 
paying a lower energy cost if the spins form a triplet, compared to the case
in which the spins form a singlet, resulting in an effective ferromagnetic
interaction proportional to the Hund exchange $J_{\text{O}}$. There are eight
contributions of this type depending on the relative order between 
processes 1 and 2,
between 3 and 4, and the two O atoms involved (above or below the Cu-Cu line).

In case (b), after three hopping processes, the holes meet at the same Cu or
O orbital and then return to the original Cu positions, or interchange them.
For the parallelogram shown in Fig. \ref{pert2}, there are 20 of such
processes which differ in the sequence of intermediate states. There is, in
addition, a factor 2 because the parallelogram can be below or above the
Cu-Cu line and an additional factor 2 from the contributions of the
parallelogram obtained from that shown in Fig. \ref{pert2} after reflection 
through a vertical plane between two Cu ions. 
For $J_{2}$, the parallelogram changes to an isosceles trapezoid,
with the base of two unit cells, 
and the
latter reflection does not change the figure, then the corresponding factor 2
is lacking and the number of different contributions is 
exactly half of those for $J_1$.

Note that the usual antiferromagnetic exchange found in the cuprates is
forbidden for a Cu-O-Cu angle of 90 degrees.

Adding carefully all contributions we obtain:

\begin{eqnarray}
J_{2} &=&4t^{2}t_{\text{O}}^{4}\left( \frac{1}{\Delta ^{4}U_{\text{Cu}}}
+\frac{4}{\Delta ^{4}\left( 2\Delta +U_{\text{O}}\right) }\right) ,  
\notag \\
J_{1} &=&2J_{2}-\frac{8t^{4}}{\Delta ^{2}}\frac{J_{\text{O}}}{\left( 2\Delta +U_{\text{O}}-2J_{\text{O}}\right) ^{2}-J_{\text{O}}^{2}},  \label{Jperturb}
\end{eqnarray}
The terms proportional to $J_{2}$ account for the antiferromagnetic
contributions and the last term in Eq. (\ref{Jperturb}) is the ferromagnetic
contribution to $J_{1}$. 

To check that all antiferromagnetic contributions have been added correctly,
we have calculated exactly the energies of a Cu$_2$O$_2$ molecule containing
the relevant orbitals of the parallelogram of Fig. \ref{pert2}, and 
obtained the correct singlet-triplet splitting for 
$t, t_{\text{O}} \rightarrow 0$ \cite{note}.

\section{Numerical results}
\label{num}

Due to the covalency of the system, the perturbative results obtained in 
the previous section for the exchange parameters $J_i$ with $i=1,2$ are not accurate 
for realistic values of the parameters of the five-band Hubbard Hamiltonian.
To derive the $J_i$ in this case, we follow an approach which has been proved
successful for example in mapping the three-band model for the cuprates 
into a spin-fermion model \cite{bat}: we assume that the effective model $H_{\text{Heis}}$ given by Eq. (\ref{eqn:heis}) retains this form and obtain the $J_i$ fitting the low lying energy levels
of the multiband Hubbard Hamiltonian. We have checked that adding an exchange
interaction to third nearest neighbors practically does not improve the fitting, which
supports the above mentioned assumption. 

Due to the rapid increase of the Hilbert space with the size of the system, 
we have done the fitting using a three-band model in a CuO chain with eight 
unit cells, eliminating one row of the original CuO$_2$ chain.
Smaller systems lead to either frustration (6 unit cells) or underestimation (4 unit cells) of the NNN interaction. While DMRG can be used to obtain the ground-state of the 5-band model with 8 unit cells, our code does not allow to obtain excited states with 
definite wave vector $k$.
According 
to perturbation theory, the results for $J_i$ should by multiplied by 
two to reach to the correct answer. 

We have calculated exactly the energies of 
the three-band model for the parameters obtained as described in section 
\ref{5b} for each wave vector and the lowest total spin projections. The sum of the squares of the differences with the corresponding 
energies of $H_{\text{Heis}}$ was minimized with respect to both $J_i$ using the Nelder-Mead procedure, starting from the perturbative regime and 
varying parameters of the multiband model in an adiabatic way to avoid 
reaching local non-physical minima. Periodic boundary conditions were used 
for $H_{\text{Heis}}$ and antiperiodic ones for the three-band model to compensate
for the fermionic sign in a state of eight particles when the last particle
is moved to the first place.

\subsection{Perturbative regime}

In order to check the regime of validity of the perturbative calculations, 
we have compared energies in this regime for different choices of parameters.

\subsubsection{Ferromagnetic contribution to $J_1$}
From Eqs. (\ref{Jperturb}), setting $t_{\text{O}}=0$ the only contribution to the perturbative calculation corresponds to the  case (a) of section \ref{pert}, which leads to a ferromagnetic NN exchange  $J_1<0$, and zero NNN exchange $J_2=0$. 

\begin{figure}[h!]
    \centering
    \includegraphics[width=0.47\textwidth]{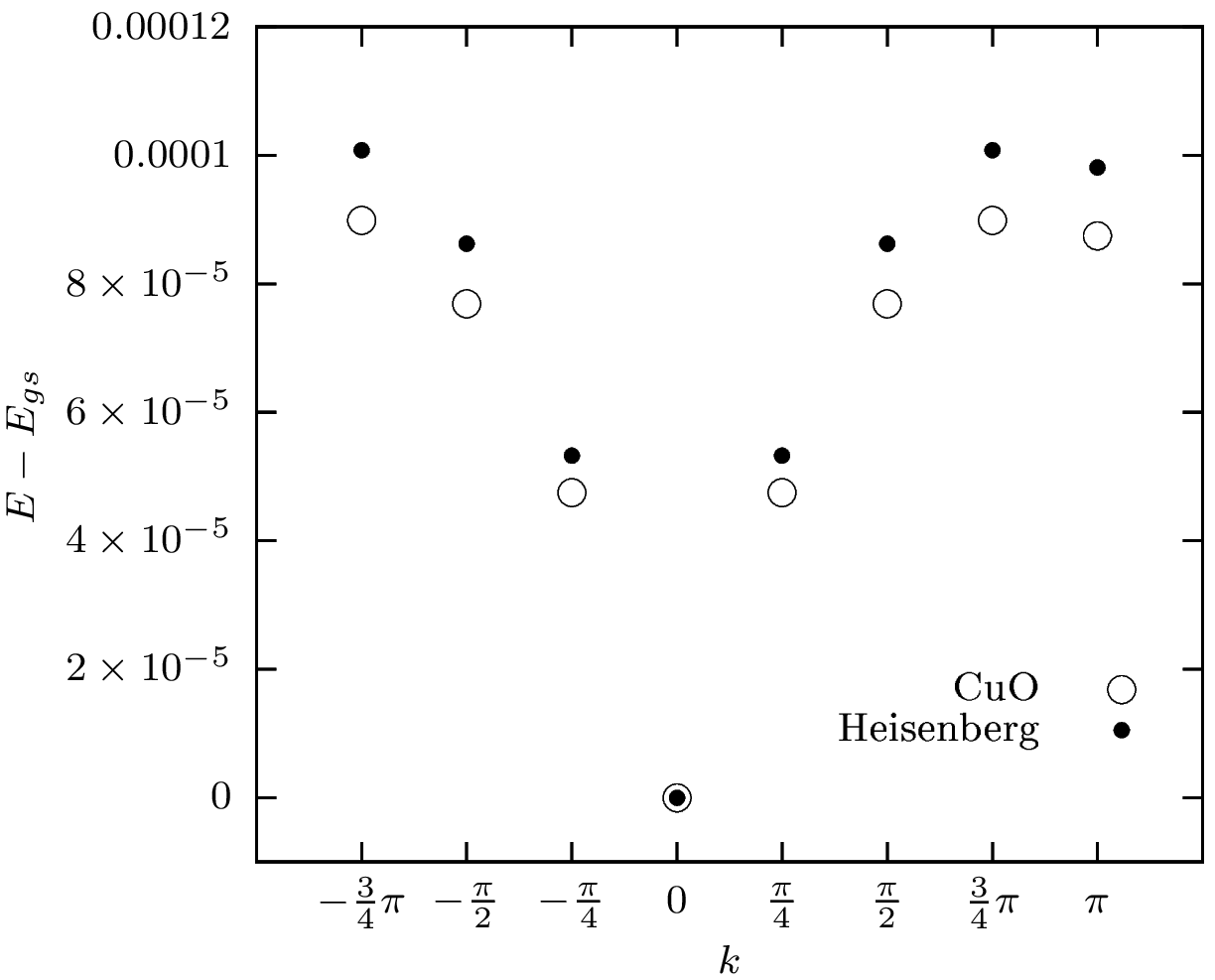}
    \caption{Energy differences with respect to the corresponding ground state energies as a function of wave vector for both models. Parameters are $t=1$eV, $t_{\text{O}}=0$, $J_{\text{O}}$=1eV, $U_{\text{O}}$=3eV, $U_{\text{Cu}}$=10eV, $\Delta$=10eV and the perturbative effective interactions: $J_1/2=-0.182$ meV $J_2=0$ and $S_z=0$}
    \label{fitJ1ferro}
\end{figure}

In Fig. \ref{fitJ1ferro} we show energies for both models for parameters of the multiband model in the perturbative regime and $t_{\text{O}}=0$. 
Here the parameters of $H_{\text{Heis}}$ are taken from 
Eqs. (\ref{Jperturb}) and not by minimization.
From these results we can see that perturbative calculations are consistent with the numerical result, with a slight difference between energies, being the energies of the effective model around $12\%$ higher. 

\subsubsection{Antiferromagnetic contributions to $J_i$}

\begin{figure}[h!]
    \includegraphics[width=0.47\textwidth]{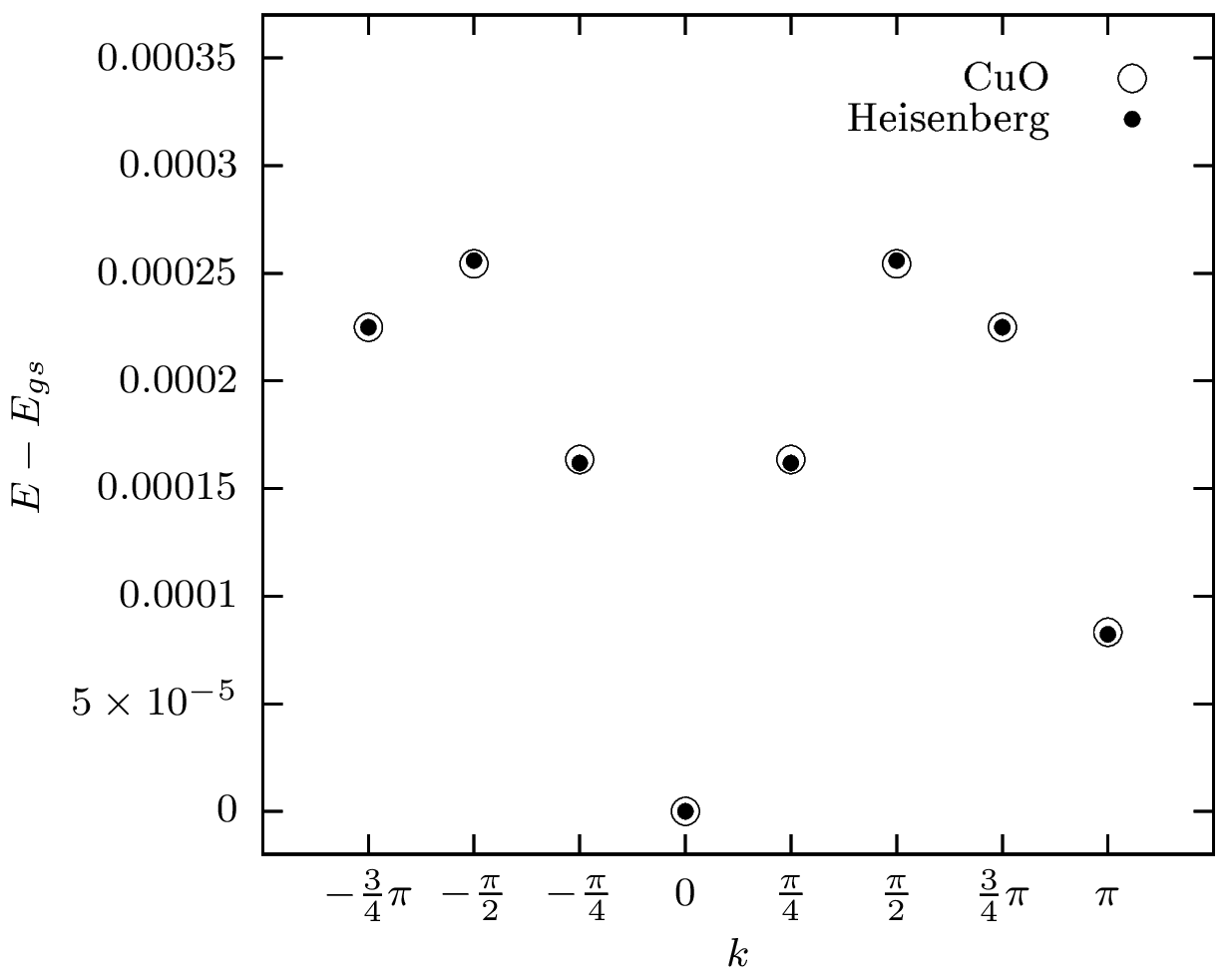}
    \caption{Energy differences with respect to the corresponding ground state energies for the multiband and the Heisenberg model for $t=1$eV, $t_{\text{O}}=1$eV, $J_{\text{O}}=0$, $U{_\text{O}}=1$eV, $U_{\text{Cu}}=100$eV, $\Delta=10$eV. The resulting
exchange parameters after the fitting procedure are $J'_1/2=0.169$ meV, $J'_2/2=0.0300$ meV}
    \label{fitJ1af}
\end{figure} 

We have also analyzed the antiferromagnetic contribution to both exchange interactions, $J_1$ and $J_2$ by considering a finite hopping between the oxygen atoms, $t_{\text{O}}$ and $J_O=0$, so that the 
ferromagnetic contributions to $J_1$ vanish. 
The energies for both models are shown in Fig. \ref{fitJ1af} for $\Delta=10$eV where we have used the optimized parameters $J_1^\prime$ and $J_2^\prime$. The Heisenberg model reproduces accurately the energies of the multiband model.

\begin{table}[h!]
\begin{tabular}{c c c c c c c c}
\hline
\hline
$\Delta$[eV] & $J_1/2$[meV] & $J_2/2$[meV] & $J_1^\prime$[meV] & $J_2^\prime$[meV] & 
$2 J_1^\prime/J_1$ & $2 J_2^\prime/J_2$ \\
\hline
10 & 0.16038 & 0.08019 & 0.16925 & 0.03004 & 
1.06 & 0.37\\
11 & 0.10049 & 0.0502 & 0.11156 & 0.02069 & 
1.11 & 0.41\\
12& 0.06559 & 0.03279 & 0.07615 & 0.0147 & 
1.16 & 0.45\\
13 & 0.0443 & 0.02215 & 0.05335 & 0.01052 & 
1.20 & 0.47\\
14 & 0.03081 & 0.01540 & 0.03831 & 0.00767 & 
1.24 & 0.50\\
15 & 0.02197 & 0.01099 & 0.02812 & 0.00573 & 
1.28 & 0.52 \\

\hline

\end{tabular}
\caption{Comparison of the exchange parameters $J_i$ that result from perturbation theory
 [see Eqs. (\ref{Jperturb})] to those obtained from a fit of the energies of
a Cu$_8$O$_8$ ring $J_i^\prime$, for several values of $\Delta$. The other parameters are 
$t=1$eV, $t_{\text{O}}=1$eV, $J_{\text{O}}=0$, $U_{\text{O}}=1$eV and $U_{\text{Cu}}=100$eV}
\label{j1afvsDelta}
\end{table}
 
To check the perturbative results for the antiferromagnetic contributions to the NN and NNN exchange parameters, we have compared the perturbative results $J_i$ with the corresponding values $J_i^\prime$ obtained from the minimization procedure
using the Cu$_8$O$_8$ ring, for different choices of $\Delta$.  
The results for the exchange parameters are shown in Table \ref{j1afvsDelta}.
The ratios $2 J_i'/J_i$ indicate how much the minimization results differ from the perturbative results of Eqs. (\ref{Jperturb}). 
We remind the reader that since the Cu$_8$O$_8$ ring has half the O atoms of the 
CuO$_2$ chain, one expects that the resulting exchange iterations are half the correct ones.
For $J_1$, in the case $\Delta=10$eV, $J_1^\prime$ is barely higher (only $0.06\%$) from the perturbative result. When $\Delta$ increases, the minimization result tends to grow around $30\%$ higher. On the other hand, the minimization results for $J_2$ tend to improve when $\Delta$ is increased, but for the case $\Delta=15$eV there is still a factor $2$ between both, being the minimization result smaller than the perturbative one. 
Nevertheless, both minimization results for NN and NNN couplings remained of the same order of magnitude of the corresponding perturbative calculation.
We believe that the discrepancy is likely due to the effect of perturbative processes of higher order (probably not contained in the Cu$_2$O$_2$ molecule mentioned at the end of Section III \cite{note}), which are still important for large values of $\Delta$. One fact that contributes to this effect is the rapid increase of the number of different 
perturbative processes of the same order involved, as the order increases. In particular, 
for the calculation of the antiferromagnetic contribution to $J_1$, the results of 80 different processes have been added.

\subsection{Non-perturbative regime}

In this subsection, we report the resulting values of $J_i$ obtained
from our fitting procedure using realistic parameters of the multiband
Hamiltonian. To this end we use the experimental determination of the charge-transfer (CTG) gap\cite{papa} of 1.95eV to fit the values of $\Delta $ in Eq. (2) for a set of realistic values for $U_{\text{O}}$. These values were obtained by extrapolating the finite-size CTGs for $L=4$ and $L=8$ cells using the five-orbital model, solved with the DMRG. In Table II we show the resulting values of $J_i^\prime$, $\Delta$ and the corresponding $\theta$ angles. These angles are the pitch angles of the spiral state which results from the frustrated spin-1/2 quantum spin chain with NN and NNN interactions reported in Ref. \onlinecite{bur}. They were extracted from Fig. 2 or Ref.  \onlinecite{bur} 
using  our results for $J_i^\prime$. 
The resulting values of the interactions are of the order of 
magnitude of those reported previously from an analysis of the 
neutron data \cite{masuda,masuda2,gip} and \textit{ab-initio} calculations \cite{gip}. 
The value of the pitch angle observed in neutron experiments is near 63 degrees \cite{park,masuda}. This points to a ferromagnetic $J_1$
and of larger magnitude than our results.

We find that for smaller values of $U_{\text{O}}$, the ferromagnetic term in Eq. (2) is favored and this leads to negative values of $J_1^\prime$. This, in turn, leads to a pitch angle $\theta < 90$ degrees since there is a larger ferromagnetic alignment between neighboring spins that competes with the antiferromagnetic NNN interaction. 
For larger values of $U_{\text{O}}$, the ferromagnetic term in Eq. (2) is weakened, leading to a larger predominance of the antiferromagnetic interaction between NN spins. 

\begin{figure}[h!]
    \includegraphics[width=0.47\textwidth]{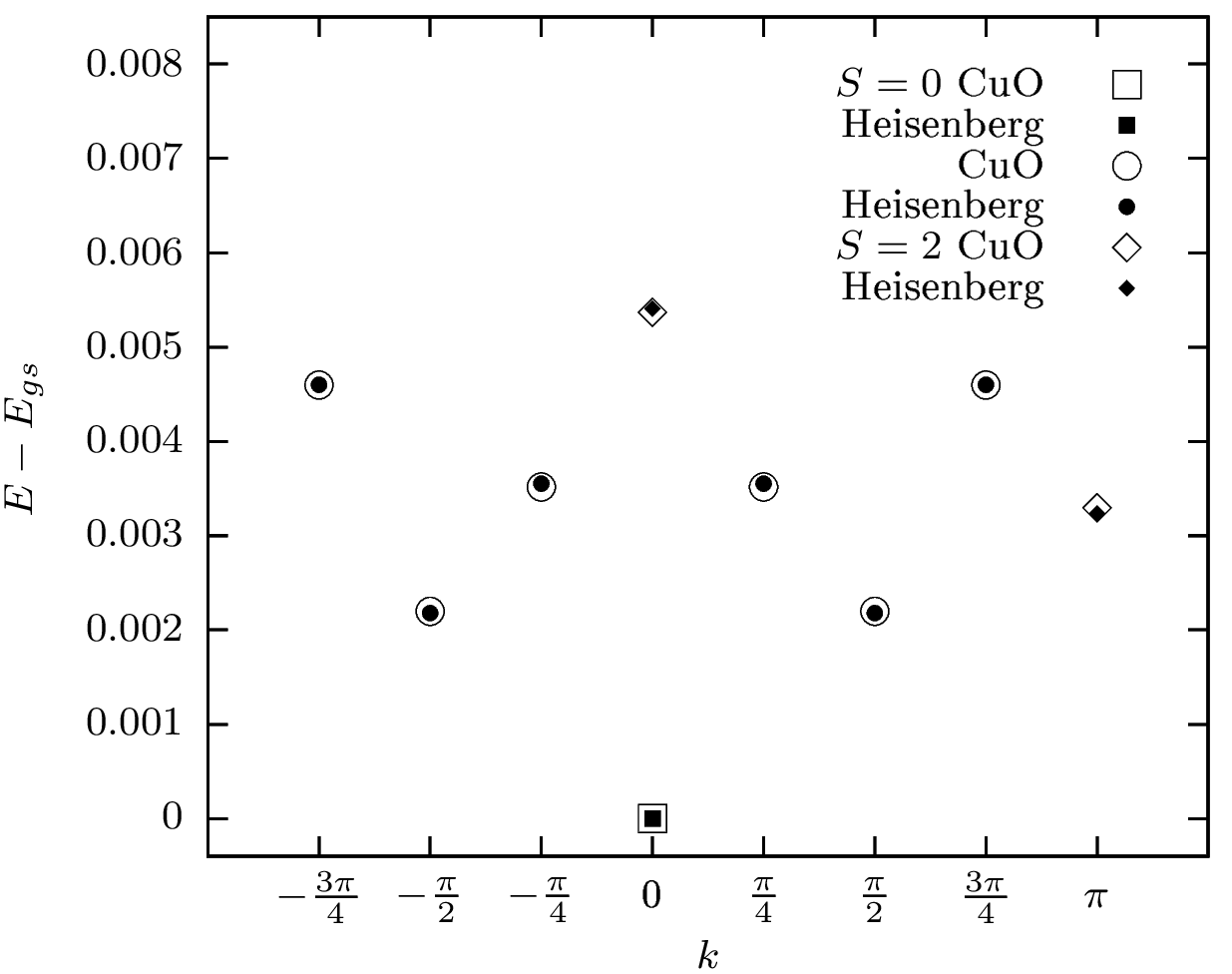}
    \caption{Energy differences with respect to the corresponding ground state energies as a function of $k$ and total spin for $t=1$eV, $t_{\text{O}}=0.36$eV, $J_{\text{O}}=0.84$eV, $U_{\text{O}}=1$eV, $U_{\text{Cu}}=10$eV, $\Delta=3.64$eV}
    \label{fit-real1}
\end{figure}

\begin{figure}[h!]
    \includegraphics[width=0.47\textwidth]{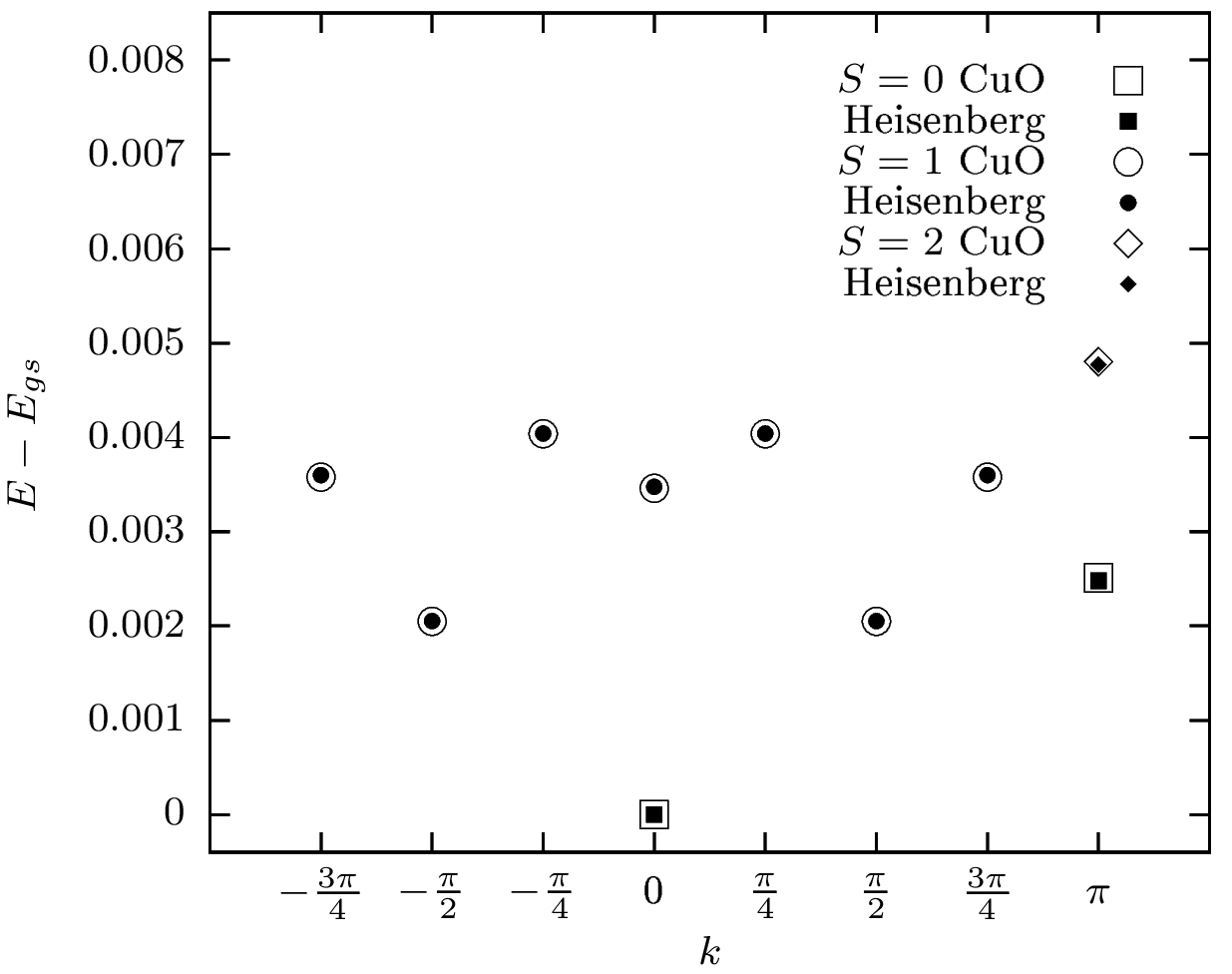}
    \caption{Same as Fig. \ref{fit-real1} for 
$U_{\text{O}}=4$eV and $\Delta=3.3$eV} 
    \label{fit-real2}
\end{figure}

A comparison of the energies in both models is shown in 
Fig. \ref{fit-real1} 
and Fig. \ref{fit-real2}, where we find an excellent correspondence between the realistic multiband Hamiltonian and the effective spin model.

\begin{table}[h!]
\begin{tabular}{c c c c c c c c c}
\hline
\hline
$U_{\text{O}}$[eV] & $J'_1$[meV] & $J'_2$[meV] & $\Delta$[eV] & $\theta$[degrees] \\
\hline
1 & 	-4.527	& 4.482	& 3.64	& 88.228 \\
2 &		-1.685	& 4.309	& 3.52	& 88.582 \\
3 &		0.577	& 4.149	& 3.42	& 90.146 \\
4 &		2.590 	& 4.140 &  3.3 & 92.48 \\

\hline
\end{tabular}
\caption{$J_i^\prime$ for a set of realistic values for $U_{\text{O}}$ and their corresponding $\Delta$ parameters as explained in the text. We also show the pitch angles $\theta $ of the resulting spin spiral state (see text)}
\label{finalJ}
\end{table}

\section{Summary and discussion}

In summary, we have derived from first principles an effective model to describe the CuO$_2$ chains in the ferroelectric material LiCu$_2$O$_2$. We began with an {\it ab-initio} calculation to determine the relevant orbitals and physical parameters from which we obtain a five-band Hubbard model containing local Coulomb and Hund interactions. Using perturbation theory in the hopping parameters we obtained the effective low-energy spin model for one hole per Cu site, leading to a one-dimensional spin-1/2 Heisenberg model with NN ($J_1$) and NNN ($J_2$) exchange interactions. These calculations shed light on the underlying physical mechanisms behind these interactions.

However, for realistic values of the parameters the system is not in the perturbative regime. Therefore, we obtained corrected values of the effective spin interactions $J_1$ and $J_2$ by fitting the lowest energy levels of a multiband Hubbard model, calculated using exact diagonalization. We rely on experimentally obtained values of the charge-transfer gap to obtain the difference 
between on-site energies of Cu and O, for which we solve the five-band Hubbard model using the Density Matrix Renormalization Group. The magnitude of $J_2$ is, in general, larger than $J_1$ and is always antiferromagnetic, while $J_1$ changes from ferro- to antiferromagnetic as the local Coulomb repulsion on the oxygen atoms increases.
The large value of $J_2$ leads to a spiral spin order, 
which is a sufficient condition for the emergence of ferroelectricity in this material, on the basis of symmetry arguments.

The pitch angle observed in neutron experiments, that has a value of around 63 degrees, is smaller than
our results. This would require larger values for $\Delta$ (smaller values of the local Coulomb interaction) which would lead to 
a more negative $J_1$ and a smaller positive $J_2$. Since a very large screening of the Coulomb interaction is unlikely, we believe that the discrepancy might be due to an underestimation of the difference between on-site energies of holes at O and Cu sites $\Delta$ as a consequence of finite-size effects. 

Nevertheless, to our knowledge, this is the first many-body calculation of the effective exchange interactions of the material.

\section*{Acknowledgments}
N. Aucar Boidi acknowledges support from ICTP through the STEP program, the ESI through the long-term program ``Tensor Networks: Mathematical Structures and Novel Algorithms'' and the ``Autumn School on Correlated Electrons:
Dynamical Mean-Field Theory of Correlated Electrons'' in which part of this work has been done.
AAA is supported by PICT 2017-2726 and AAA and KH by PICT 2018-01546 of the ANPCyT, and by PIP 2015-0538 of CONICET, Argentina.



\begin{references}
\bibitem{note} Specifically taking $t=1$eV, $t_{\text{O}}=0.1$eV, $U_{\text{O}}=J_{\text{O}}=0$, $U_{\text{Cu}} \rightarrow \infty$, and $\Delta=10$eV, the perturbative result
is 7\% larger than the result for the Cu$_2$O$_2$ molecule. For larger $\Delta=50$eV, 
the difference falls below 1\%.
\end{references}
\end{document}